\documentclass[letterpaper,twocolumn,english,superscriptaddress,letterpaper, nofootinbib]{revtex4}

\usepackage[]{fontenc}
\usepackage[latin9]{inputenc}
\usepackage{textcomp}
\usepackage{amsmath}
\usepackage{graphicx}
\usepackage{amssymb}

\makeatletter

\pdfpageheight\paperheight
\pdfpagewidth\paperwidth

\newcommand{\lyxmathsym}[1]{\ifmmode\begingroup\def\b@ld{bold}
  \text{\ifx\math@version\b@ld\bfseries\fi#1}\endgroup\else#1\fi}

\@ifundefined{textcolor}{}
{%
 \definecolor{BLACK}{gray}{0}
 \definecolor{WHITE}{gray}{1}
 \definecolor{RED}{rgb}{1,0,0}
 \definecolor{GREEN}{rgb}{0,1,0}
 \definecolor{BLUE}{rgb}{0,0,1}
 \definecolor{CYAN}{cmyk}{1,0,0,0}
 \definecolor{MAGENTA}{cmyk}{0,1,0,0}
 \definecolor{YELLOW}{cmyk}{0,0,1,0}
 }

\makeatother

\usepackage{babel}

\begin{document}

\title{Quantum many-body interactions in digital oxide superlattices}

\author{Eric J. Monkman \footnote{These authors contributed equally to this work.}}

\affiliation{Laboratory of Atomic and Solid State Physics, Department of Physics,
Cornell University, Ithaca, New York 14853, USA}

\author{Carolina Adamo \footnotemark[\value{footnote}]}

\affiliation{Department of Materials Science and Engineering, Cornell University,
Ithaca, New York 14853, USA}

\author{Julia A. Mundy}

\affiliation{School of Applied and Engineering Physics,
Cornell University, Ithaca, New York 14853, USA}

\author{Daniel E. Shai}

\author{John W. Harter}

\author{Dawei Shen}

\affiliation{Laboratory of Atomic and Solid State Physics, Department of Physics,
Cornell University, Ithaca, New York 14853, USA}

\author{Bulat Burganov}

\affiliation{Laboratory of Atomic and Solid State Physics, Department of Physics,
Cornell University, Ithaca, New York 14853, USA}

\author{David A. Muller}

\affiliation{School of Applied and Engineering Physics,
Cornell University, Ithaca, New York 14853, USA}

\affiliation{Kavli Institute at Cornell for Nanoscale Science, Ithaca, New York
14853, USA}

\author{Darrell G. Schlom}

\affiliation{Department of Materials Science and Engineering, Cornell University,
Ithaca, New York 14853, USA}

\affiliation{Kavli Institute at Cornell for Nanoscale Science, Ithaca, New York
14853, USA}

\author{Kyle M. Shen}

\email[email: ]{kmshen@cornell.edu}

\affiliation{Laboratory of Atomic and Solid State Physics, Department of Physics,
Cornell University, Ithaca, New York 14853, USA}

\affiliation{Kavli Institute at Cornell for Nanoscale Science, Ithaca, New York
14853, USA}

\date[Received 13 March 2012 \textbar Accepted 19 July 2012 \textbar Published Online 19 August 2012 \textbar DOI:10.1038/nmat3405]{}

\maketitle

\textbf{Controlling the electronic properties of interfaces has enormous scientific and technological implications and has been recently extended from semiconductors to complex oxides which host emergent ground states not present in the parent materials \cite{Chakhalian06,Logvenov09,Ohtomo04,Okamoto04,Jang11}. These oxide interfaces present a fundamentally new opportunity where, instead of conventional bandgap engineering, the electronic and magnetic properties can be optimized by engineering quantum many-body interactions \cite{Jang11,Dagotto07,Chakhalian11}. We utilize an integrated oxide molecular-beam epitaxy and angle-resolved photoemission spectroscopy system to synthesize and investigate the electronic structure of superlattices of the Mott insulator LaMnO$_{3}$ and the band insulator SrMnO$_{3}$. By digitally varying the separation between interfaces in (LaMnO$_{3}$)$_{2n}$/(SrMnO$_{3}$)$_{n}$ superlattices with atomic-layer precision, we demonstrate that quantum many-body interactions are enhanced, driving the electronic states from a ferromagnetic polaronic metal to a pseudogapped insulating ground state. This work demonstrates how many-body interactions can be engineered at correlated oxide interfaces, an important prerequisite to exploiting such effects in novel electronics.}

Exotic magnetic phases \cite{Chakhalian06}, high-T$_{c}$ superconductivity \cite{Logvenov09}, and two-dimensional correlated electron systems \cite{Ohtomo04,Okamoto04,Jang11} are only a few examples of novel states recently realized at complex oxide interfaces. While the electronic properties of conventional semiconductor heterostructures can be described by one-electron theories, performing such calculations for correlated materials is far more challenging due to competing many-body interactions. Understanding these correlated interfaces has been complicated by the inability to probe their underlying electronic structure and quantum many-body interactions \cite{Dagotto07,Chakhalian11}, necessitating the development of new advanced spectroscopic probes. 

Here we employ a combination of oxide molecular-beam epitaxy (MBE) and angle-resolved photoemission spectroscopy (ARPES) using an integrated system to first engineer and then investigate digital superlattices of (LaMnO$_{3}$)$_{2n}$/(SrMnO$_{3}$)$_{n}$. Due to the surface sensitivity of ARPES, it has not yet been possible to measure atomically pristine oxide interfaces. Our integrated system circumvents this problem by allowing synthesis and measurement within the same ultra-high vacuum manifold, avoiding any surface contamination. The manganites present an ideal case for modifying electronic and magnetic properties through interfacial engineering due to their competing interactions and wide variety of ground states \cite{Dagotto01}. Bulk LaMnO$_{3}$ and SrMnO$_{3}$ are antiferromagnetic Mott and band insulators, respectively, and La$_{2/3}$Sr$_{1/3}$MnO$_{3}$ is a ferromagnetic metal which exhibits colossal magnetoresistance around its Curie temperature of 370 K. Increasing the separation between the LaMnO$_{3}$ and SrMnO$_{3}$ layers with integer $n$ has been shown to drive a crossover from a ferromagnetic metallic ($n < 3$) to ferromagnetic insulating ground state ($n \ge 3$) \cite{Salvador99,Bhattacharya08,Adamo09} whose origin is currently not understood, as theoretical studies predict metallic interfaces for large $n$ \cite{Dong08,Nanda09}. 

The tunability of oxide heterostructures can arise from either controlling band alignments or structural potentials as is achieved in conventional semiconductors, or by taking advantage of the strong many-body interactions that are uniquely accessible in correlated materials. We utilize ARPES to reveal that while the band structure remains largely unchanged with interfacial separation, a large pseudogap is opened within 800 meV of $E_{F}$ for $n = 3$, and closes either upon warming into the paramagnetic state or reducing the interfacial separation ($n \le 2$). Our work provides the first direct observation of how quantum many-body interactions can be engineered in artificial materials constructed with atomic-layer precision to control the electronic ground state.

%%%%%%%%%%%%%%%%%%%%
\begin{figure*}
\includegraphics{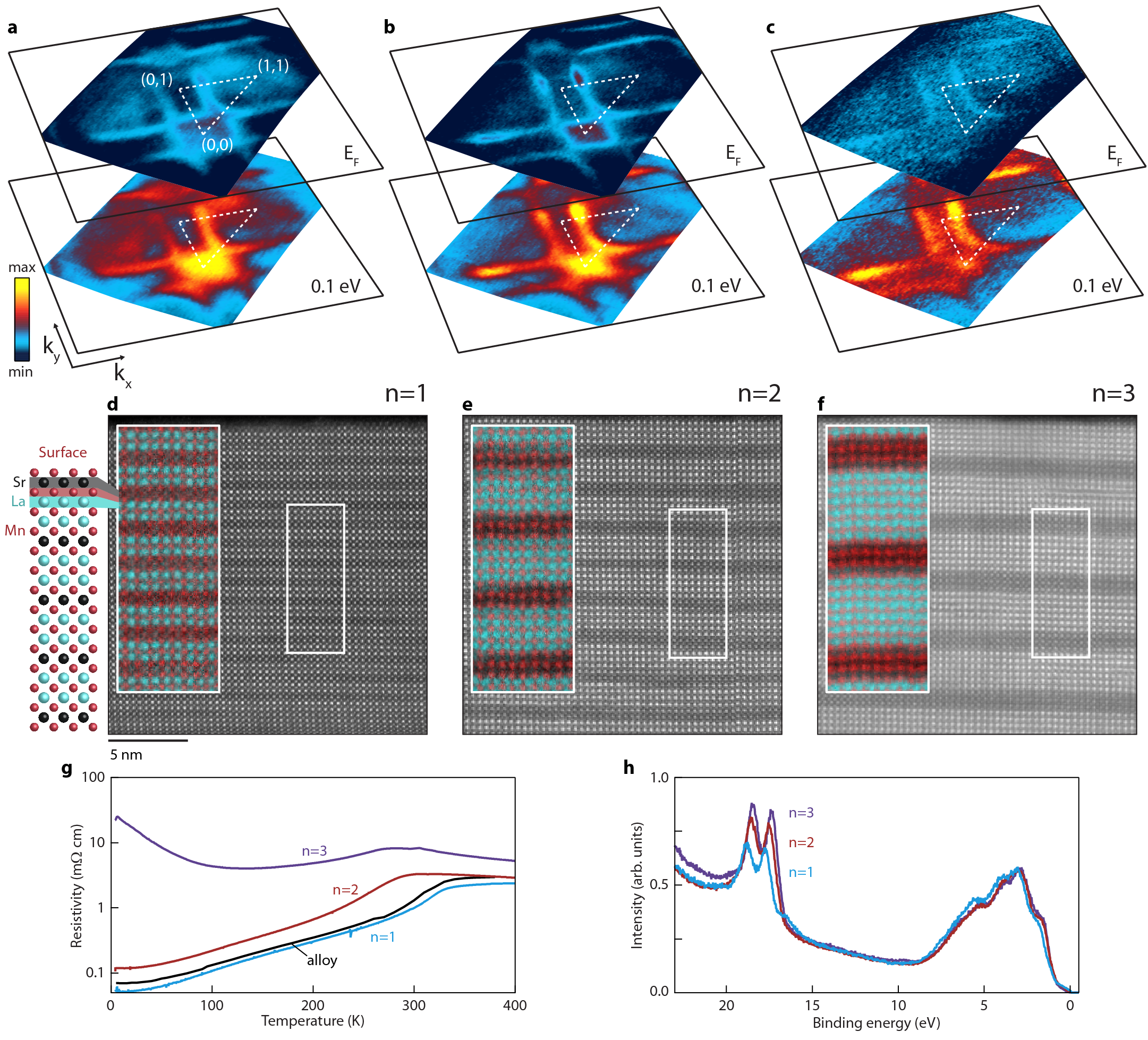}
\caption{\textbf{Overview of the superlattices' electronic structure and properties.} \textbf{a-c,}~$\mathbf{k}$-resolved maps of photoemission spectral weight for (LaMnO$_{3}$)$_{2n}$/(SrMnO$_{3}$)$_{n}$ with $n$ = 1, 2, and 3 respectively, at $T = 10$ K. Maps are at $E_{F}$ and 0.1 eV binding energy as indicated. \textbf{d-f,} High angle annular dark field scanning transmission electron micrographs of $n$ = 1, 2, and 3 samples that were measured by ARPES, along with a schematic of the $n = 1$ structure with MnO$_{2}$ surface termination. Enlarged electron energy loss spectroscopic images (insets) show La in turquoise and Mn in red, and demonstrate well-ordered films with clear separation of Sr and La. Distortions in the EELS map are not structural, but are artifacts from sample drift during data acquisition. Images over a very wide field of view can be found in the supplementary information. \textbf{g,} Resistivity of superlattices as a function of $n$ and temperature showing the crossover to insulating behavior at $n=3$ (data from Ref \cite{Adamo09}). Also shown for comparison, the resistivity of a random alloy La$_{2/3}$Sr$_{1/3}$MnO$_{3}$ film grown under identical conditions. \textbf{h,} The valence bands of the three superlattices are nearly identical over a wide energy range, except for the increasing intensity of the Sr $4p$ core states around 18 eV due to the termination of each superlattice by $n$ unit cells of SrMnO$_{3}$.
\label{fig:3FSs}}

\end{figure*}
%%%%%%%%%%%%%%%%%%%%

In Fig. \ref{fig:3FSs}, we show $\mathbf{k}$-resolved spectral weight maps for the $n$ = 1, 2, and 3 samples of (LaMnO$_{3}$)$_{2n}$/(SrMnO$_{3}$)$_{n}$. All films were terminated with $n$ layers of SrMnO$_{3}$ with a MnO$_{2}$ surface (Fig. \ref{fig:3FSs}d), and are expected to be non-polar due to their inversion symmetric structure. In Fig. \ref{fig:3FSs}a and b, the Fermi surfaces (FSs) of the metallic $n$ = 1 and 2 materials are apparent and consist of two Mn $e_{g}$ derived states: a hole pocket of predominantly $d_{x^{2} - y^{2}}$ character around the Brillouin zone (BZ) corner, and a smaller electron pocket of primarily $d_{3z^{2} - r^{2}}$ character around the zone center. For the insulating $n$ = 3 sample,  spectral weight at $E_{F}$ is suppressed, although clear states are still observed below $E_{F}$. In Fig. \ref{fig:3FSs}d, e, and f we show elementally resolved scanning transmission electron micrographs of the same samples measured by ARPES demonstrating atomically abrupt LaO/MnO$_{2}$/SrO and SrO/MnO$_{2}$/LaO interfaces. In Fig. \ref{fig:3FSs}g, we show resistivity for $n$ = 1, 2, and 3 superlattices grown using the same approach \cite{Adamo09}, showing the metal-insulator crossover for $n \ge 3$. In Fig. \ref{fig:3FSs}h, the valence bands of $n$ = 1, 2, and 3 are shown.

%%%%%%%%%%%%%%%%%%%%
\begin{figure}
\includegraphics{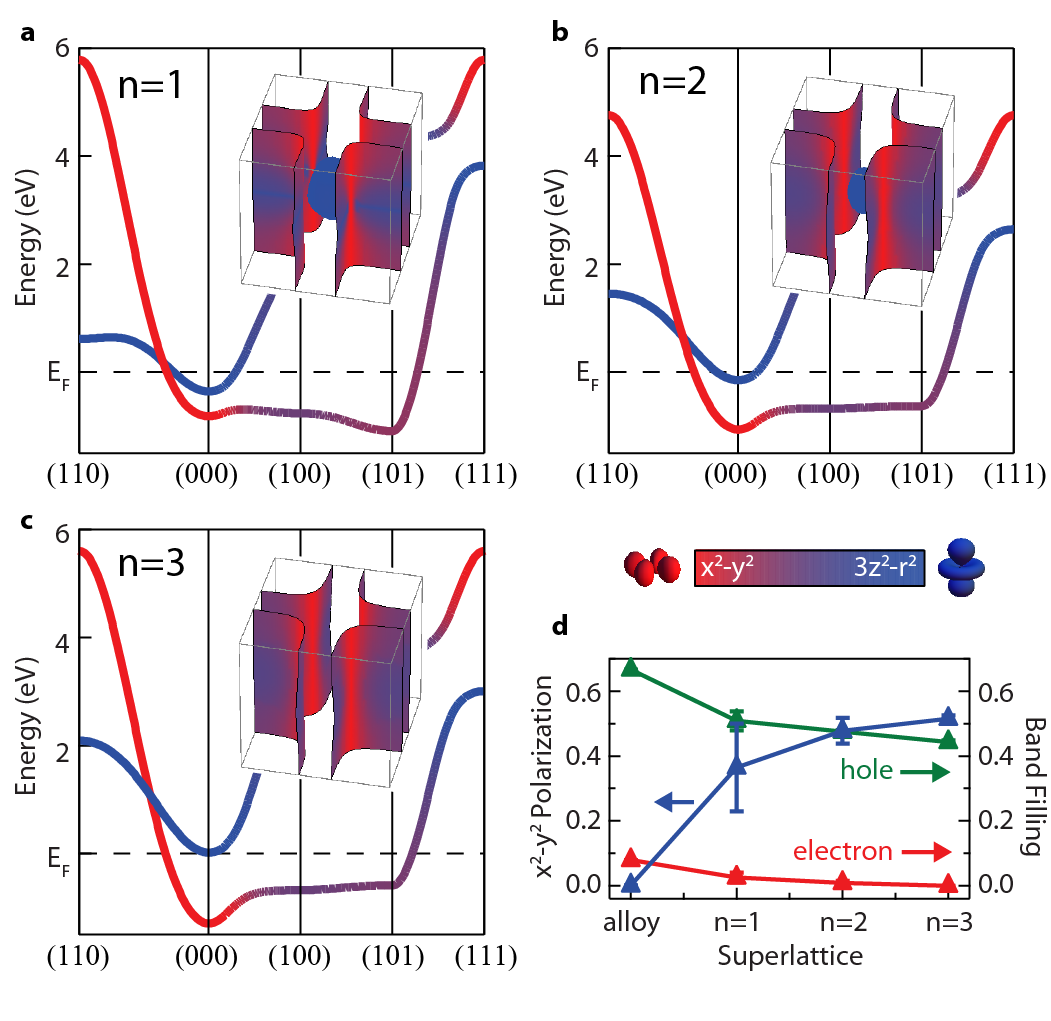}
\caption{\textbf{The tight-binding parametrization.} \textbf{a-c,}~TB bandstructures and FSs extracted from our ARPES data for the $n$ = 1, 2, and 3 superlattices. Orbital character throughout the BZ is indicated by each band's color. \textbf{d,} Orbital polarization and filling of the electron and hole pockets from the TB model for the three superlattices and the random alloy. Error bars are determined from the maximum and minimum estimated size of the electron pocket from our ARPES data, which dominates the uncertainty of the TB model (supplementary information).
\label{fig:TB}}

\end{figure}
%%%%%%%%%%%%%%%%%%%%

%%%%%%%%%%%%%%%%%%%%
\begin{figure*}
\includegraphics{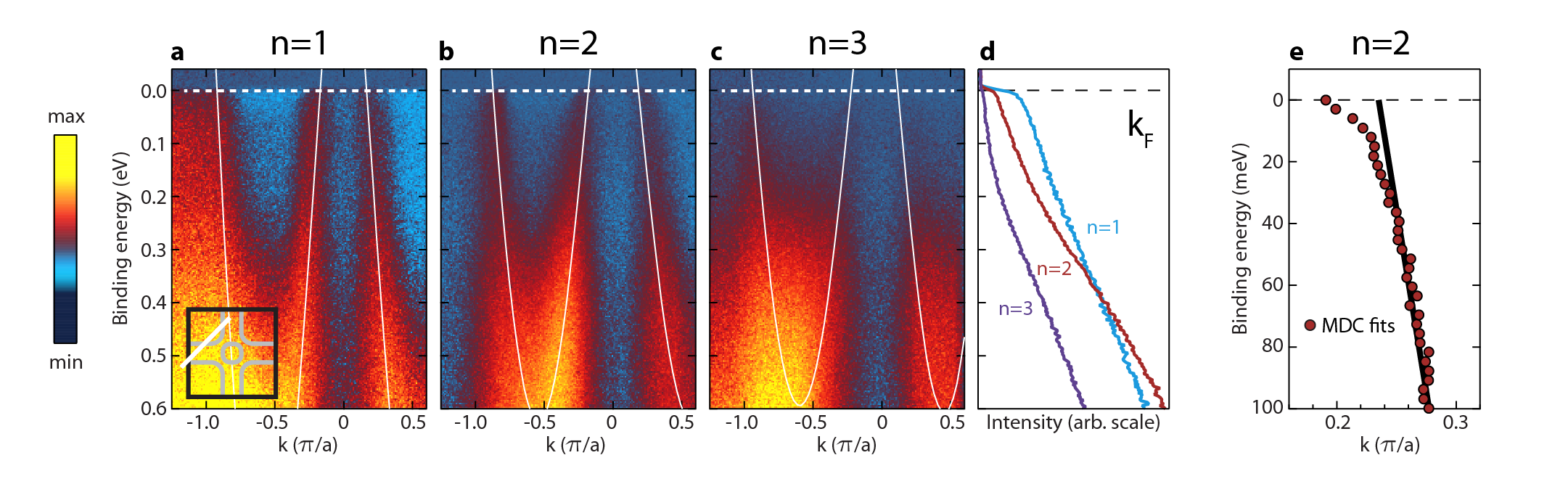}
\caption{\textbf{Electronic structure of the hole pockets.} \textbf{a-c,}~ARPES spectra along the $\mathbf{k}$-path illustrated by the inset of panel a, showing the hole-pocket band crossing $E_{F}$ at three points. TB fits are overlaid in white as guides to the eye. A non-dispersive background has been subtracted from the ARPES data to more clearly illustrate the bandstructure (supplementary information). \textbf{d,} Energy-dependent photoemission intensity (EDCs) at $\mathbf{k_{F}}$ of the hole pocket. \textbf{e,} ARPES band dispersion for n=2 compared to the linear extrapolation of the dispersion for $E>0.075$ eV, showing a kink at 35 meV.
\label{fig:Bands}}

\end{figure*}
%%%%%%%%%%%%%%%%%%%%

%%%%%%%%%%%%%%%%%%%%
\begin{figure}
\includegraphics{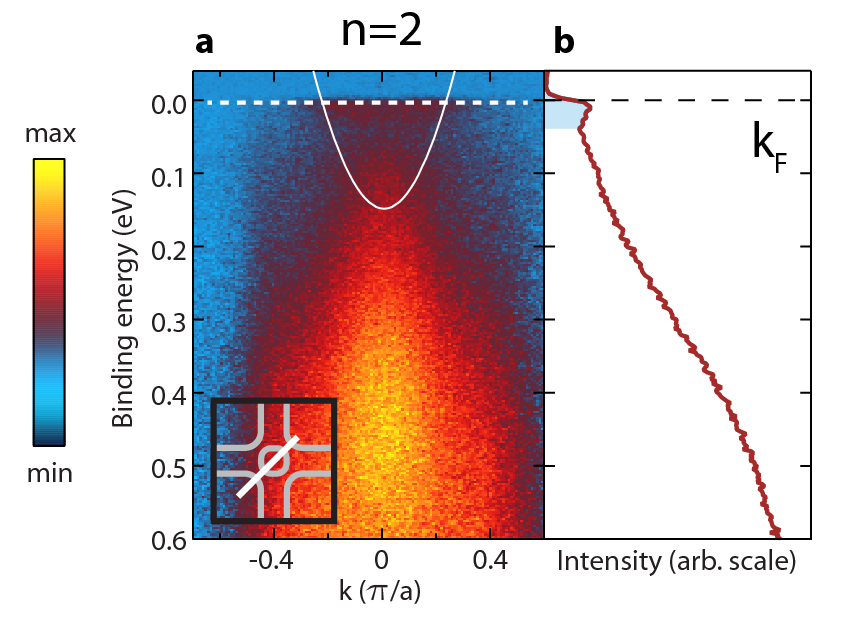}
\caption{\textbf{Strongly renormalized QP peak.} \textbf{a,}~ARPES spectra along the $\mathbf{k}$-path illustrated by the inset of panel a for the $n$ = 2 superlattice. The TB band for the electron pocket is shown by the white line as a guide to the eye. A non-dispersive background has been subtracted from the ARPES data to more clearly illustrate the bandstructure (supplementary information). \textbf{b,} Energy-dependent photoemission intensity (EDC) at $\mathbf{k_{F}}$ of the electron pocket. The QP peak is schematically illustrated by the blue shaded area.
\label{fig:Bands_epocket}}

\end{figure}
%%%%%%%%%%%%%%%%%%%%

We observe that the electron pocket decreases in size as $n$ increases in Fig. \ref{fig:3FSs}a-c, indicating the preferential filling of $d_{x^{2} - y^{2}}$ orbitals, suggesting an approach towards an interfacial 2D electronic structure. To quantify these changes, we fit the measured FSs and dispersions to a simple tight-binding (TB) parametrization (supplementary information), shown in Fig. \ref{fig:TB} and overlaid onto data in Figs. \ref{fig:Bands} and \ref{fig:Bands_epocket}. Estimating the carrier density from counting the Luttinger volume of the FSs gives higher hole concentrations ($x \approx 0.50 \pm 0.1$) than  would be expected from a random alloy ($x = 0.33$), as expected from a LaMnO$_{3}$-SrMnO$_{3}$ interface. From our TB wavefunctions, we can also estimate the orbital polarization defined as $(N_{(x^{2}-y^{2})}-N_{(3z^{2}-r^{2})})/(N_{(x^{2}-y^{2})}+N_{(3z^{2}-r^{2})})$, where $N$ is the integral of the partial density of states up to $E_{F}$. The orbital polarization increases from 0\% for cubic La$_{2/3}$Sr$_{1/3}$MnO$_{3}$ to approximately 50\% for $n = 3$, consistent with x-ray absorption measurements that find $d_{x^{2} - y^{2}}$ polarization at the interfaces \cite{Aruta09}. This polarization is dominated by the $d_{x^{2} - y^{2}}$ character of the hole-like sheets, although our orbital polarization never approaches 100\% because their TB wavefunctions still retain non-negligible $d_{3z^{2} - r^{2}}$ character. Despite its simplicity, our TB model should qualitatively describe the change in orbital polarization with $n$, although more sophisticated density functional calculations would be necessary to obtain more accurate wavefunctions. 

In Figs. \ref{fig:Bands} and \ref{fig:Bands_epocket}, we show the quasiparticle (QP) dispersion in energy versus momentum along cuts shown in the insets of Figs. \ref{fig:Bands}a and \ref{fig:Bands_epocket}a. The $n$ = 1 and 2 samples exhibit well-defined and dispersive bands (Fig. \ref{fig:Bands}a,b). A sharp QP peak can only be observed for $n$ = 2 (Fig. \ref{fig:Bands_epocket}) and, due to photoelectron final state effects \cite{Krempasky08}, can be attributed to the interfacial states increasing confinement to 2D with $n$. The peak-dip-hump structure, where the coherent QP peak is dominated by a broad hump of incoherent spectral weight, is a signature of correlated systems and has been observed in the cuprates \cite{RMP03} and other manganites \cite{Mannella05}. Fig. \ref{fig:Bands}e shows a kink in the dispersion for the $n$ = 2 $d_{x^2  - y^2}$ band within 35 meV of $E_{F}$. The ratio of band velocities at high and low energy gives $v_{F, \mathrm{high}} / v_{F, \mathrm{low}} = 3.7 \pm 0.6$. Within a weak-coupling scenario, this would correspond to a mass renormalization $m^{\ast}/m_{\mathrm{band}}$ = 3.7, although this falls well into the strong coupling regime. Similar features have been observed in other correlated systems which exhibit strong electron-boson interactions, such as the cuprates \cite{RMP03}, and some bilayer manganites, where a similar velocity renormalization was observed at nearly the same energy ($v_{F, \mathrm{high}} / v_{F, \mathrm{low}} = 5.6$) and was attributed to strong electron-phonon coupling \cite{Mannella05}. 

Unlike the metallic superlattices, $n$ = 3 exhibits only pseudogapped intensity at $E_{F}$ (Fig. \ref{fig:Bands}d), similar to polaronic systems with strong electron-phonon coupling\cite{Chuang01,Massee11}. Although the total spectral weight is generally conserved, the weight at low-energies is pushed to higher energy scales which may then be obscured by the valence band. Despite the pseudogap, the $n$ = 3 sample still exhibits the underlying hole-like $d_{x^{2} - y^{2}}$ dispersion at higher binding energies with a comparable bandwidth and similar remnant FS to the $n$ = 1 and 2 samples, as shown in Fig. \ref{fig:3FSs}c. 

%%%%%%%%%%%%%%%%%%%%
\begin{figure}
\includegraphics{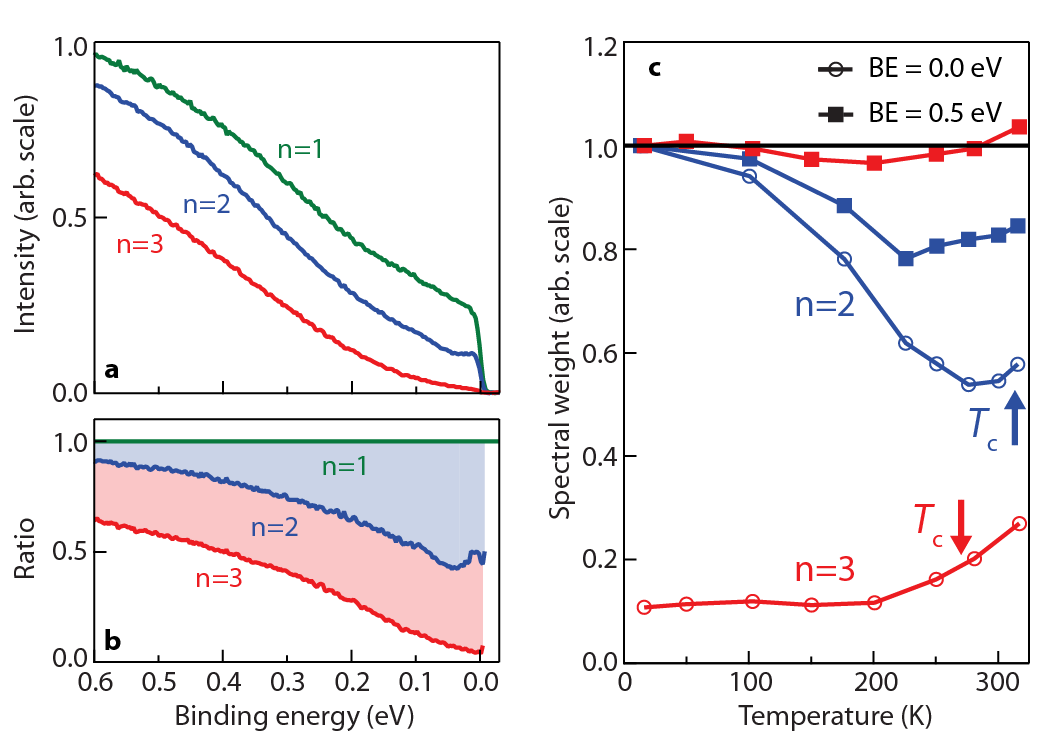}
\caption{\textbf{The pseudogap and temperature-dependent spectral weight.} \textbf{a,}~The angle-integrated spectral weight for the $n$ = 1, 2, and 3 superlattices, showing pseudogap behavior in the $n=3$ film. \textbf{b,} The data from panel a normalized to the spectral weight of the $n=1$ superlattice, highlighting the strong energy-dependence of the pseudogap. \textbf{c,} Temperature dependence of the spectral weight at the BZ center for the $n$ = 2 and 3 superlattices. Open circles show the spectral weight within 50 meV of $E_{F}$, closed squares show the spectral weight at a binding energy of 400 to 550 meV. Also indicated in c are the Curie temperatures ($T_{c}$) of each material. Data in c are normalized to unity at 20 K, except for the $n = 3$ open circles. Here, $n = 3$ data are normalized such that the integrated weight over $E_{F}$ to 8.4 eV is equal to the $n = 2$ integral over the same window, allowing for a meaningful comparison between the two films.
\label{fig:PG}}

\end{figure}
%%%%%%%%%%%%%%%%%%%%

The similar dispersion of $n$ = 3 to $n$ = 1 and 2 demonstrates that the insulating behavior is not caused by the opening of a conventional bandgap as might be the case in semiconductor interfaces, and is consistent with recent resonant scattering measurements \cite{Smadici07}. This suggests the pseudogap is the origin of the $n$-driven metal-insulator crossover, also  supported by our temperature dependent measurements (Fig. \ref{fig:PG}c). The metallic $n$ = 2 superlattice shows a reduction in  weight near $E_{F}$ with increasing temperature, consistent with a loss of coherent QPs in the paramagnetic state. On the other hand,  the spectral weight of $n$ = 3 increases above T$_{c}$ as the pseudogap  fills in, consistent with its resistivity approaching the metallic superlattices in the paramagnetic state.

We have a variety of reasons that indicate our ARPES measurements of the topmost interfaces are representative of the bulk properties. First, we observe states at $E_{F}$ for the metallic $n = 1$ and 2 superlattices, while the insulating $n$ = 3 superlattice exhibits a pseudogap. Second, we observe a correspondence between the temperature dependence of our spectra and the bulk Curie temperature. Third, measurements of SrMnO$_{3}$ and LaMnO$_{3}$ films (supplementary information) do not exhibit any of the near-$E_{F}$ electronic structure of the superlattices which thus must arise from the LaMnO$_{3}$-SrMnO$_{3}$ interface. Fourth, FS volumes give hole concentrations close to $x$ = 0.50 as might be expected from the LaMnO$_{3}$-SrMnO$_{3}$ interface, as opposed to $x = 0.33$. Fifth, the near-$E_{F}$ suppression of spectral weight for $n$ = 3 vs $n$ = 1 is highly energy dependent close to $E_{F}$ (Fig. \ref{fig:PG}a,b), suggesting that this effect should not be due to surface sensitivity, since $\lambda_{mfp}$ is effectively energy-independent in such a narrow range. Finally, our LEED and RHEED patterns show $2 \times 4$ and $3 \times 3$ reconstructions associated with the SrMnO$_{3}$ termination (supplementary information), but the ARPES data do not exhibit any evidence of such a periodicity, again suggesting that the near-$E_{F}$ states arise from the buried interface.

It has been suggested that reduced dimensionality could drive the $n \ge 3$ superlattices insulating via Anderson localization \cite{Dong08}, where a Coulomb gap could form due to interactions between localized electrons. Despite their suppressed intensity, we observe well-defined bands for $n$ = 3, demonstrating states that are delocalized over more than $8\times8$ unit cells ($\Delta k <1/8$ of BZ), outside of the conventional scenario for the Coulomb gap \cite{CG75} or Anderson localization \cite{Edwards78}. Our density of states near $E_{F}$ (which follows $\approx\omega^{2}$) also deviates from the linear dependence expected for a 2D Coulomb gap, while TEM measurements demonstrate a nearly disorder-free structure. Furthermore, a change in the hole concentration would not explain the crossover at higher $n$, since La$_{1-x}$Sr$_{x}$MnO$_{3}$ has no ferromagnetic insulating state at large $x$.

It is then natural to consider the quantum many-body interactions that are inherent to the manganites as the origin of the metal-insulator crossover observed with $n$. These interactions are known to give rise to insulating ordered states, as the effective dimensionality (i.e. coupling along the $c$-axis) is lowered in the Ruddlesden-Popper series of manganites, (La,Sr)$_{m+1}$Mn$_{m}$O$_{3m+1}$, where $m$ is the number of MnO$_{2}$ planes per unit cell \cite{Moritomo96}. Bilayer $m$ = 2 La$_{2-2x}$Sr$_{1+2x}$Mn$_{2}$O$_{7}$ is a pseudogapped ferromagnetic ``bad metal'' with a low temperature resistivity two orders of magnitude higher than the metallic 3D perovskite ($m$ = $\infty$). The quasi-two-dimensional $m$ = 1 compound La$_{1-x}$Sr$_{1+x}$MnO$_{4}$ is insulating for all Sr concentrations \cite{Moritomo96,Dagotto01} due to the formation of charge, spin, or orbital order \cite{Larochelle05,Moritomo96}, and exhibits fully gapped spectral weight at $E_{F}$ and a remnant Fermi surface observed by ARPES in La$_{0.5}$Sr$_{1.5}$MnO$_{4}$ \cite{Evtushinsky10}.

Our measurements demonstrate that dimensionality also plays a similar role in the superlattices, as the partially occupied interfacial states become progressively separated and 2D with increasing $n$ as the effective hopping between interfaces is reduced. The similarities between their spectral features indicate the many-body interactions responsible for the properties of the single and bilayer manganites are also paramount for the superlattices. The small QP weight and kink observed in the $n = 2$ superlattice suggests a metallic state comprised of coherent polarons which are strongly coupled to the lattice, orbital, and/or magnetic degrees of freedom which reduce the QP residue \cite{Chuang01,Mannella05,RMP03}. Recent calculations also suggest that electron-lattice coupling should strongly influence the properties of manganite superlattices and interfaces  \cite{Lin08,Iorio11}. Reducing the dimensionality from $n$ = 2 to 3 results in a situation where the lowered dimensionality and possibly increased nesting in 2D may enhance quantum fluctuations towards the insulating charge, spin, and orbitally ordered states, such as those observed in the single layer $m$ = 1 manganites \cite{Chuang01,Trinckauf12}. These quantum fluctuations can disrupt the coherence of the fragile polaronic metallic state, giving rise to the weakly insulating / bad metal state observed in the bilayer manganites, as proposed by Massee \emph{et al.} and Salafranca \emph{et al.} \cite{Massee11, Salafranca09}, thereby resulting in the pseudogap and loss of coherent QP weight in the $n = 3$ superlattice.

Our measurements of (LaMnO$_{3}$)$_{2n}$/(SrMnO$_{3}$)$_{n}$ demonstrate how the interplay of interactions and dimensionality can be used to control the properties of correlated oxide interfaces. By decoupling the LaMnO$_{3}$-SrMnO$_{3}$ interfaces, we are able to reduce the effective dimensionality, driving the polaronic metal at small $n$ into a pseudogapped insulator. Demonstrating both the control and understanding of the interactions at these correlated interfaces should be a key step towards the rational manipulation and optimization of their functionality for potential applications.

\section*{Methods}

(LaMnO$_{3}$)$_{2n}$/(SrMnO$_{3}$)$_{n}$ films of 20-25 nm thickness were grown on SrTiO$_{3}$ substrates by MBE in two different systems: a dual chamber Veeco 930 MBE and a dual chamber Veeco GEN10 MBE, both equipped with reflection high-energy electron diffraction. Immediately after growth, they were transferred through ultrahigh vacuum to our ARPES chamber in under 300 s. All films were terminated with $n$ layers of SrMnO$_{3}$. Further details on growth and characterization can be found in the supplementary information.

ARPES measurements were performed with a VG Scienta R4000 electron analyzer and a VUV5000 helium plasma discharge lamp and monochromator using 40.8 eV photons. The base pressure of the ARPES system was 4$\times$10$^{-11}$ Torr, and data were taken at below 20 K unless specified otherwise. Constant-energy maps (Fig. 1) consist of $2\times10^{4}$ spectra integrated within $\pm$30 meV of the specified energy and taken with an energy resolution of 40 meV. Measurements of the QP dispersion, as presented in Figs. 3 and 4, were taken with a resolution of 10 meV.

Electron energy loss spectroscopic imaging (EELS-SI) and high angle annular dark field scanning transmission electron microscopy (HAADF-STEM) images were recorded from cross sectional specimens in the 100 keV NION UltraSTEM. The Mn concentration is the integrated Mn-L$_{2,3}$ edge, the La concentration is the integrated M$_{4,5}$.

\begin{acknowledgments}
We thank L. Fitting Kourkoutis for helpful discussions and E. Kirkland for technical assistance. This work was supported by the National Science Foundation (DMR-0847385), the MRSEC program through DMR-1120296, IMR-0417392, and DMR-9977547 (Cornell Center for Materials
Research), a Research Corporation Cottrell Scholars award (20025), and NYSTAR. J.A.M. acknowledges support from the A.R.O. in the form of a NDSEG fellowship. E.J.M. acknowledges NSERC for PGS support. 
\end{acknowledgments}

\section*{Author contributions}
ARPES data was collected by E.J.M., D.E.S., J.W.H., D.S., B.B., and K.M.S.
and analyzed by E.J.M. and K.M.S. Film growth and x-ray diffraction were performed by C.A. Electron microscopy and spectroscopy measurements were performed by J.A.M. and D.A.M. The manuscript was prepared by E.J.M. and K.M.S. D.G.S. and K.M.S. planned and supervised the study. All authors discussed results and commented on the manuscript.

\end{document}

\title{Quantum many-body interactions in digital oxide superlattices}

\author{Eric J. Monkman \footnote{These authors contributed equally to this work.}}

\affiliation{Laboratory of Atomic and Solid State Physics, Department of Physics,
Cornell University, Ithaca, New York 14853, USA}

\author{Carolina Adamo \footnotemark[\value{footnote}]}

\affiliation{Department of Materials Science and Engineering, Cornell University,
Ithaca, New York 14853, USA}

\author{Julia A. Mundy}

\affiliation{School of Applied and Engineering Physics,
Cornell University, Ithaca, New York 14853, USA}

\author{Daniel E. Shai}

\author{John W. Harter}

\author{Dawei Shen}

\affiliation{Laboratory of Atomic and Solid State Physics, Department of Physics,
Cornell University, Ithaca, New York 14853, USA}

\author{Bulat Burganov}

\affiliation{Laboratory of Atomic and Solid State Physics, Department of Physics,
Cornell University, Ithaca, New York 14853, USA}

\author{David A. Muller}

\affiliation{School of Applied and Engineering Physics,
Cornell University, Ithaca, New York 14853, USA}

\affiliation{Kavli Institute at Cornell for Nanoscale Science, Ithaca, New York
14853, USA}

\author{Darrell G. Schlom}

\affiliation{Department of Materials Science and Engineering, Cornell University,
Ithaca, New York 14853, USA}

\affiliation{Kavli Institute at Cornell for Nanoscale Science, Ithaca, New York
14853, USA}

\author{Kyle M. Shen}

\email[Author to whom correspondence should be addressed: ]{kmshen@cornell.edu}

\affiliation{Laboratory of Atomic and Solid State Physics, Department of Physics,
Cornell University, Ithaca, New York 14853, USA}

\affiliation{Kavli Institute at Cornell for Nanoscale Science, Ithaca, New York
14853, USA}

\maketitle
\tableofcontents

\section*{Growth technique and structural characterization}
\begin{figure}
\includegraphics{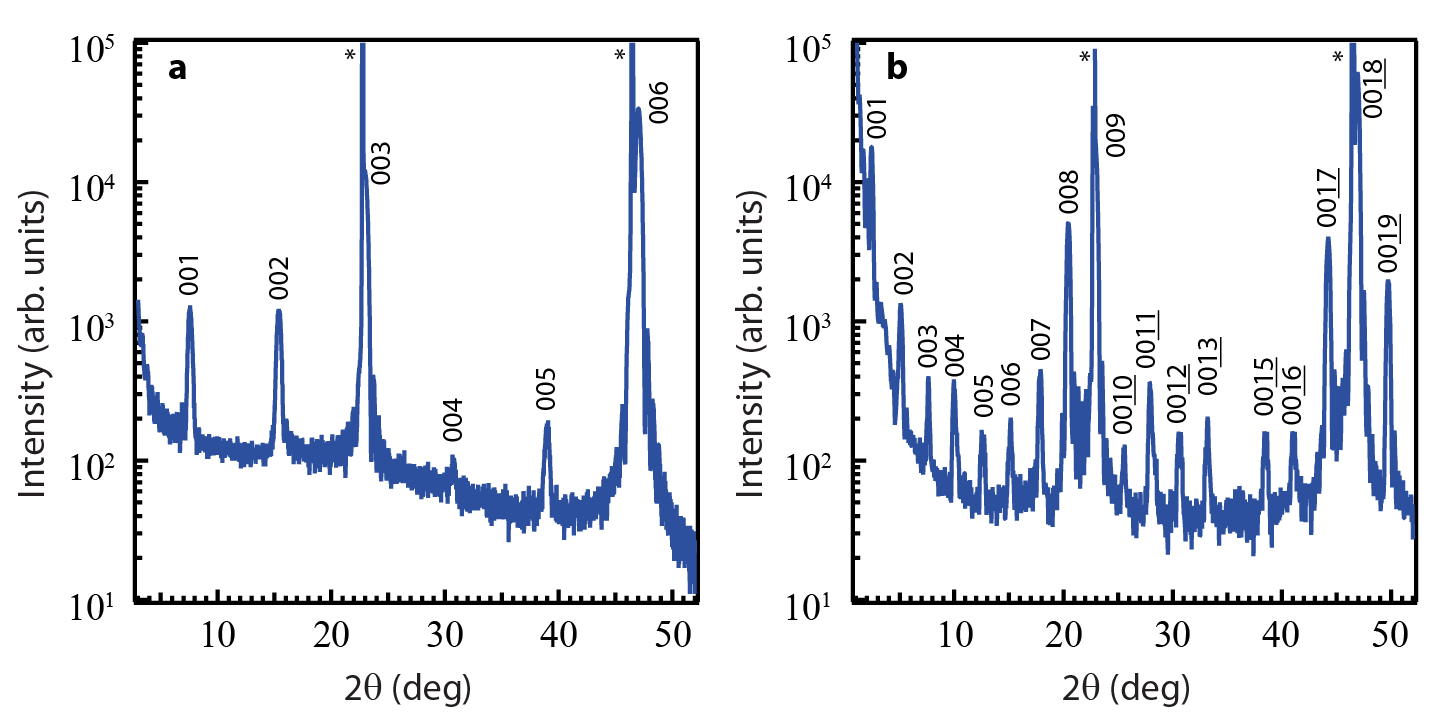}\caption{X-ray diffraction curves of the (\textbf{a}) $n$ = 1 and (\textbf{b}) $n$ = 3 superlattices measured by ARPES. Diffraction peak indices are indicated for the films; substrate peaks are denoted by the *.  \label{fig:X-ray}}
\end{figure}

Superlattices were grown using shuttered layer-by-layer deposition \cite{Haeni00} on buffered-HF treated (100)-SrTiO$_{3}$ substrates \cite{Koster98} in a reactive molecular-beam epitaxy system equipped with reflection high-energy electron diffraction (RHEED) \cite{Adamo08}. A substrate temperature of 750 $^{\circ}$C and an oxidant (O$_{2}$+10\% O$_{3}$) background partial pressure of $5\times10^{-7}$ Torr, which was kept constant until the temperature of the substrate dropped below 250 $^{\circ}$C, were used. All films measured in this study were 20 nm to 25 nm thick, and were terminated with $n$ layers of SrMnO$_{3}$, where a layer corresponds to a formula-unit-thick layer along the growth direction. On reaching 250 $^{\circ}$C, samples were immediately transferred in ultra-high vacuum ($\approx10^{-10}$ Torr) to the ARPES cryostat and cooled. X-ray diffraction data for the $n = 1$ and $n = 3$ superlattices measured by ARPES are shown in Fig. S\ref{fig:X-ray}.

In order to allow ARPES to probe the buried interface, our films were terminated with (SrMnO$_{3}$)$_{n}$ rather than the thicker (LaMnO$_{3}$)$_{2n}$ layer. To avoid surface effects arising from the polarity of the LaMnO$_{3}$ layers, our thin films were also made to be inversion symmetric by initiating growth on the SrTiO$_{3}$ substrates with SrMnO$_{3}$ layers. This introduces a very slight change of the global doping of the entire film by at most $\Delta x \le 0.03$ away from $x = 1/3$.

\begin{figure}
\includegraphics{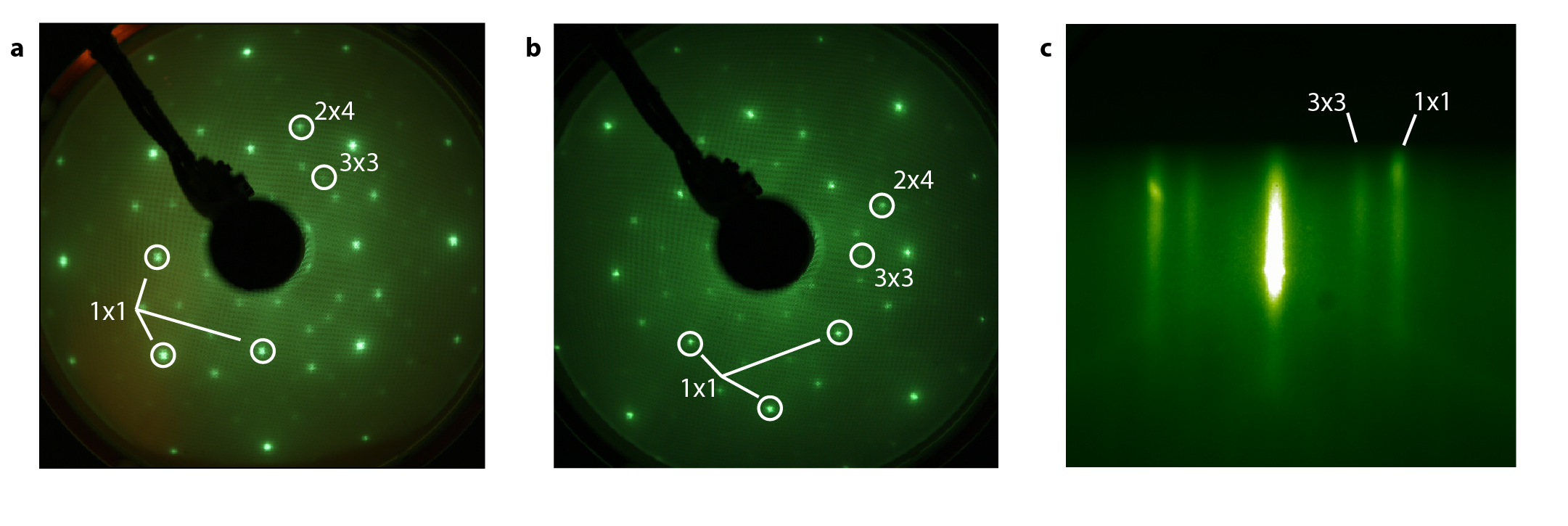}\caption{LEED pattern taken with 100 eV electrons from an $n$ = 3 superlattice (\textbf{a}) and a SrMnO$_{3}$ film (\textbf{b}), both with MnO$_{2}$ surfaces. Diffraction peaks corresponding to the unreconstructed surface and two reconstructions are indicated. Note that the SrMnO$_{3}$ LEED pattern is rotated by 45 degrees. \textbf{c,} RHEED pattern from the MnO$_{2}$ surface of the same $n$ = 3 film during growth. \label{fig:LEED}}
\end{figure}

The high structural quality of the film surface was verified after ARPES measurements with low energy electron diffraction (LEED). In Fig. S\ref{fig:LEED}a we present a LEED image taken from an $n$ = 3 superlattice after remaining in the ARPES chamber for 8 days. We observe sharp diffraction peaks, a $2 \times 4$ surface reconstruction, and a $3 \times 3$ surface reconstruction also seen by RHEED during growth (Fig. S\ref{fig:LEED}c). This demonstrates the high crystallinity of the surface of our films, and proves that the pristine surface from growth is maintained throughout the transfer to our ARPES chamber and subsequent measurement. The origin of these surface reconstructions are not yet entirely understood, but we find that they are generic to the MnO$_{2}$ surface of the perovskite manganites, and not unique to our superlattices. Both the $2 \times 4$ and $3 \times 3$ reconstructions are observed on undoped SrMnO$_{3}$ films with MnO$_{2}$ surface termination (Fig. S\ref{fig:LEED}b). The $3 \times 3$ reconstruction is also routinely observed on the MnO$_{2}$ surface by reflection high-energy electron diffraction (RHEED) during MBE growth of both the superlattices and La$_{1-x}$Sr$_{x}$MnO$_{3}$ films (e.g. Fig. S\ref{fig:LEED}c). We also observe the $2 \times 4$ surface reconstruction on La$_{0.7}$Sr$_{0.3}$MnO$_{3}$ films, along with other groups who find the same reconstruction on La$_{0.6}$Sr$_{0.4}$MnO$_{3}$ films grown by pulsed-laser deposition \cite{Horiba03}. As mentioned in the main text, we do not observe signatures of these reconstructions in our ARPES data, and since they are present for both insulating and metallic films of widely varying composition and structure, they cannot be responsible for the spectra that we report. On the other hand, we do observe a weak $c(2 \times 2)$ reconstruction in our Fermi surface maps. This reconstruction is expected to exist throughout the superlattice due to its stability in LaMnO$_{3}$ \cite{Nanda09}, and so we do not attribute it to being solely a surface effect.

\begin{figure}
\includegraphics{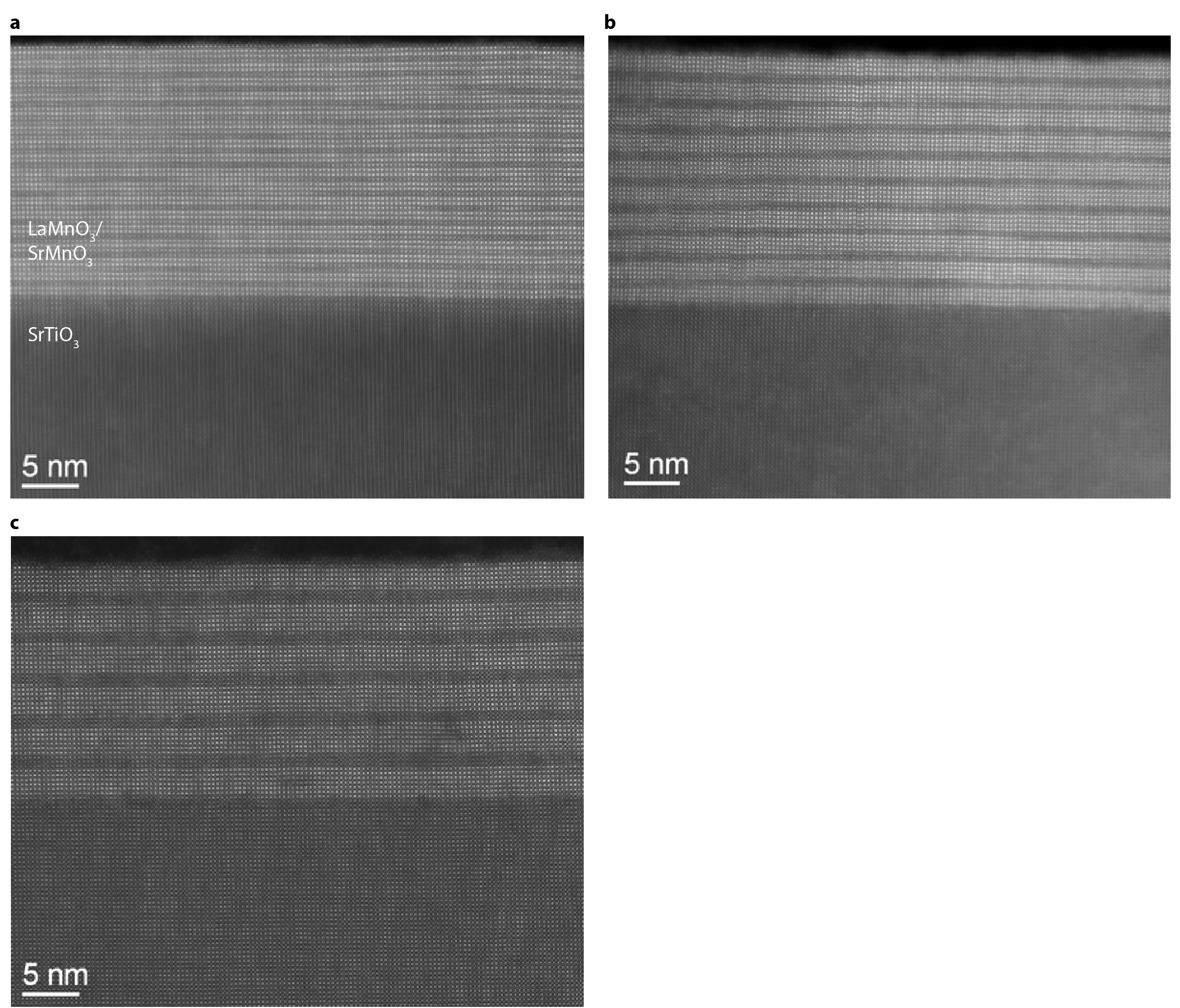}\caption{HAADF-STEM images of the same  (\textbf{a}) $n = 1$, (\textbf{b}) $n = 2$, and (\textbf{c}) $n = 3$ (LaMnO$_{3}$)$_{2n}$/(SrMnO$_{3}$)$_{n}$/SrTiO$_{3}$ films measured by ARPES.  The films show a coherent interface between the film and the substrate free of observable defects and a clear repetition of the LaMnO$_{3}$ and SrMnO$_{3}$ layering to form the desired superlattices. \label{fig:TEM}}
\end{figure}

\begin{figure}[!htb]
\includegraphics{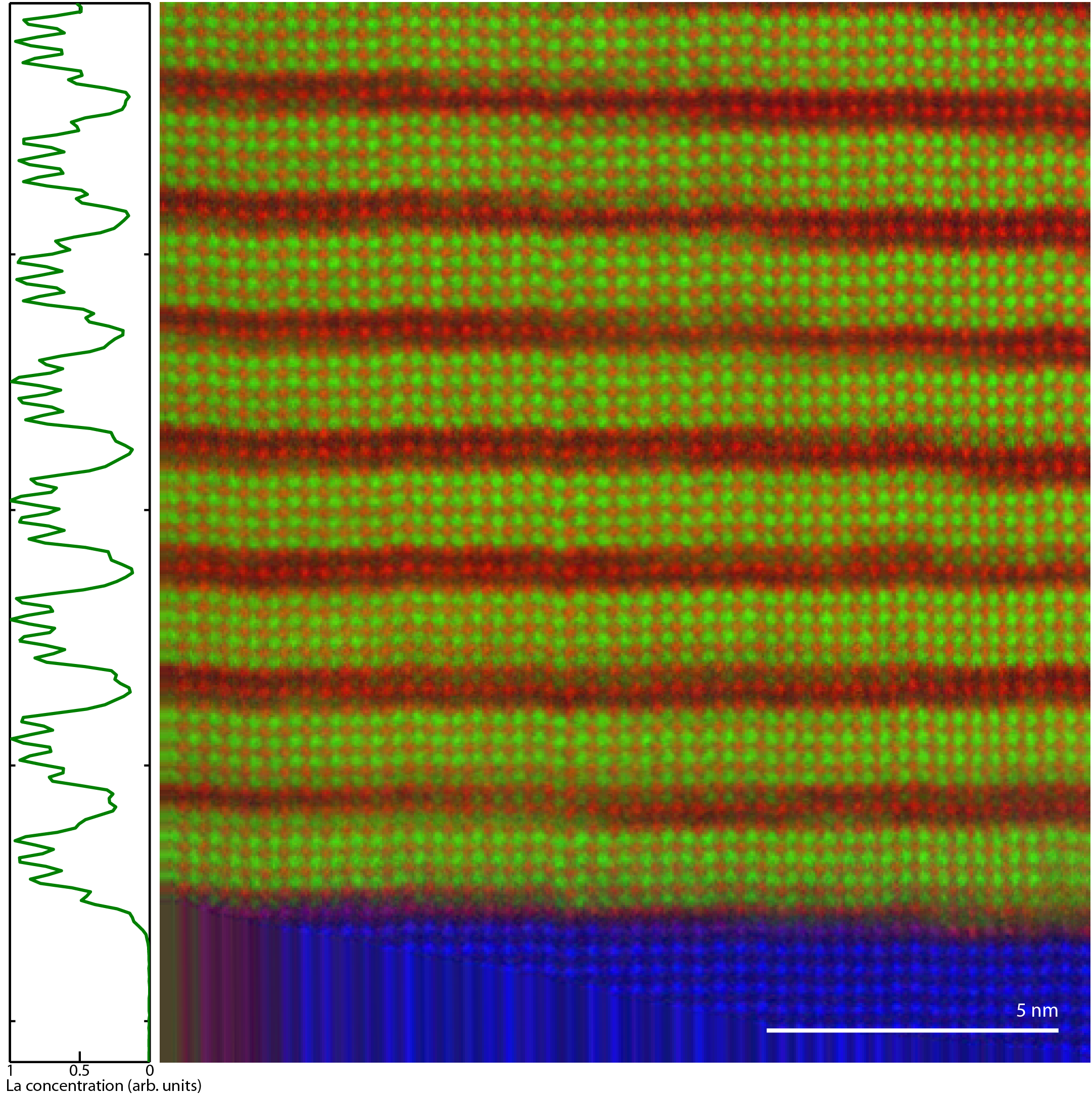}\caption{EELS map over a wide field of view from an $n = 2$ (LaMnO$_{3}$)$_{2n}$/(SrMnO$_{3}$)$_{n}$/SrTiO$_{3}$ film measured by ARPES, showing La in green, Mn in red, and Ti in blue. Steps in the LaMnO$_{3}$/SrMnO$_{3}$ interfaces follow the terraces of the SrTiO$_{3}$ substrate. Left:  the La concentration along the growth direction of the film (obtained by integrating the La-M$_{4,5}$ intensity across the image) showing sharp interfaces between LaMnO$_{3}$ and SrMnO$_{3}$ lacking any systematic asymmetry. Streaks in the bottom left corner of the EELS map are an artifact of post-acquisition drift correction. \label{fig:EELSn2}}
\end{figure}

\begin{figure}
\includegraphics{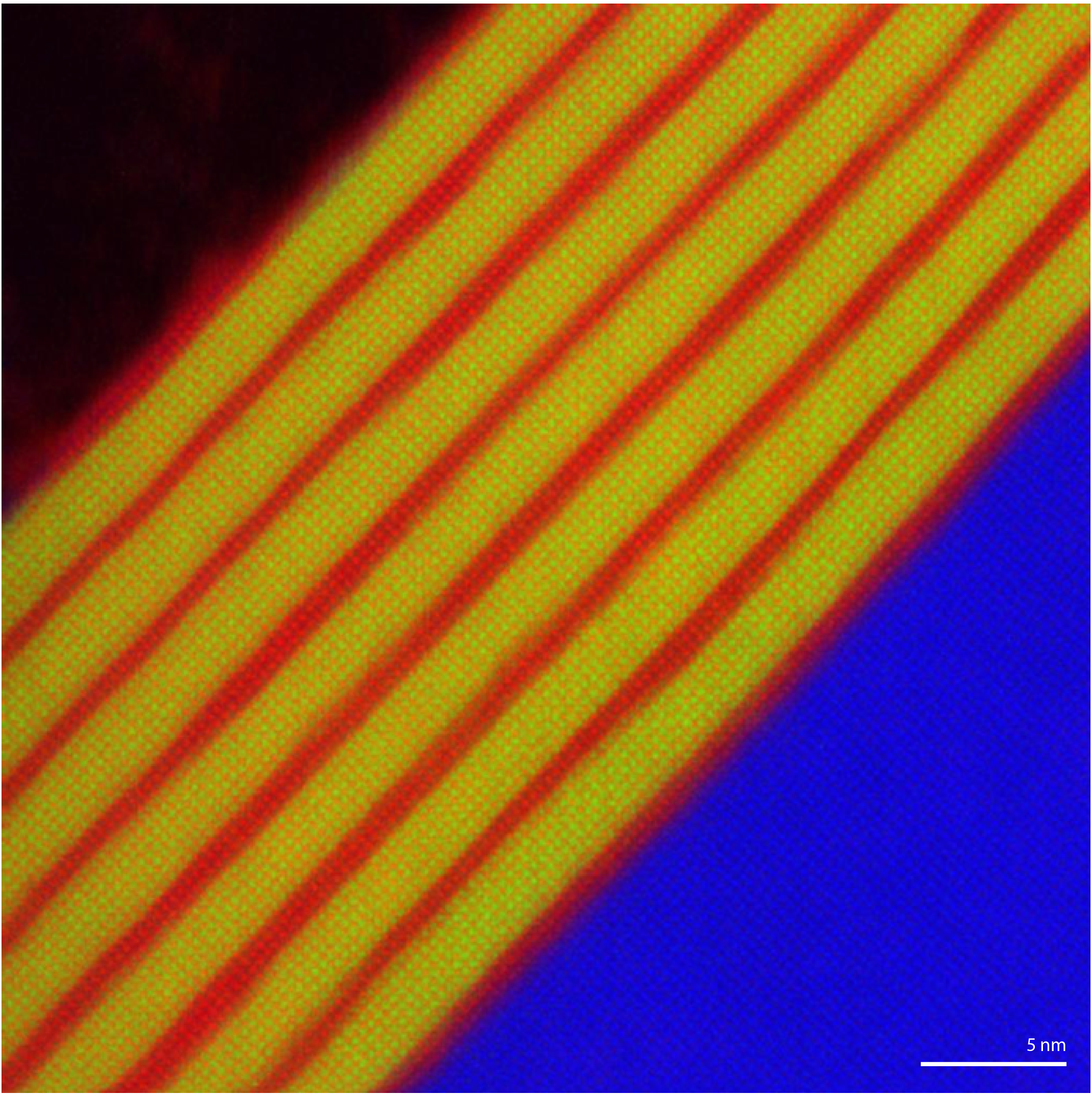}\caption{EELS map over a wide field of view from an $n = 3$ (LaMnO$_{3}$)$_{2n}$/(SrMnO$_{3}$)$_{n}$/SrTiO$_{3}$ film measured by ARPES, showing La in green, Mn in red, and Ti in blue. The irregularity of the topmost surface in this image is an artifact of the preparation procedure for EELS and HAADF-STEM measurements, and does not reflect the topmost surface of the as-grown film. \label{fig:EELSn3}}
\end{figure}

Electron energy loss spectroscopic imaging (EELS-SI), recorded from cross sectional specimens in the 100 keV NION UltraSTEM, was used to investigate three of the films measured by ARPES. The Mn concentration is the integrated Mn-L$_{2,3}$ edge, the La concentration is the integrated La-M$_{4,5}$, and the Ti concentration is the integrated Ti-L$_{2,3}$. As shown in Fig. 1d-f, Fig. S\ref{fig:TEM}, Fig. S\ref{fig:EELSn2}, and Fig. S\ref{fig:EELSn3} all samples show a clear repetition of the LaMnO$_{3}$ and SrMnO$_{3}$ layers corresponding to the $n$ = 1, 2, and 3 layering patterns. The high angle annular dark field scanning transmission electron microscopy (HAADF-STEM) images show a coherent interface between the film and substrate, free of defects (Fig. S\ref{fig:TEM}). An apparent slight modulation of the interfaces observable in EELS images is an artifact of sample drift during acquisition, and is absent in the more quickly acquired HAADF-STEM images. May et al. \cite{May08} found a strong structural asymmetry between LaMnO$_{3}$/SrMnO$_{3}$ and SrMnO$_{3}$/LaMnO$_{3}$ interfaces in (LaMnO$_{3}$)$_{11.8}$/(SrMnO$_{3}$)$_{4.4}$ superlattices, which was found to significantly effect the superlattice's magnetic properties. We note that in our extensive EELS investigations, we observed no signatures of such an asymmetry for $n = 1$ and $n = 2$ superlattices (Fig. S\ref{fig:EELSn2}), and only a very weak asymmetry for $n = 3$ when examined over very wide regions (Fig. S\ref{fig:EELSn3}). Thus, we do not expect that the asymmetry reported in ref. \cite{May08} adversely effects the properties of the films reported in our study. Although we do not understand the difference between our samples and those of May et al., the asymmetric roughening trend would be consistent with a Stranski-Krastanov growth mode for the LaMnO$_{3}$ layer, with the onset for island formation somewhere between 6 (ours) and 11 layers (May's). 

\section*{Tight-binding parametrization}
The (LaMnO$_{3}$)$_{2n}$/(SrMnO$_{3}$)$_{n}$ superlattice contains 3$n$ inequivalent Mn sites, and a full tight-binding parametrization would contain 6$n$ $e_{g}$ orbitals and many free parameters. In the interest of using the simplest possible model to represent our data, and noting that we only wish to parametrize the bandstructure at the interface, we expect a model containing one $d_{3z^{2}-r^{2}}$ and one
$d_{x^{2}-y^{2}}$ state to be adequate. Thus, we use a model defined by \cite{Ederer07};

\begin{align*}
t_{\pm a\hat{x}}  =   \frac{t_{1}}{4}  \left(\begin{array}{cc} 1 & -\sqrt{3}\\ -\sqrt{3} & 3\end{array} \right)
 & \hspace{1.9cm}
t_{\pm a\hat{y}}  =   \frac{t_{1}}{4}  \left(\begin{array}{cc} 1 & \sqrt{3}\\ \sqrt{3} & 3\end{array}\right)
& t_{\pm a\hat{z}}=   \alpha t_{1}\left(\begin{array}{cc} 1 & 0\\ 0 & 0\end{array}\right)
\\
\\
t_{\pm a\hat{x}\pm a\hat{y}}=\frac{t_{2}}{2}\left(\begin{array}{cc}
1 & 0\\
0 & -3\end{array}\right) & \hspace{1.9cm}t_{\pm a\hat{x}\pm a\hat{z}}=\frac{\alpha t_{2}}{2}\left(\begin{array}{cc}
-2 & \sqrt{3}\\
\sqrt{3} & 0\end{array}\right) & t_{\pm a\hat{y}\pm a\hat{z}}=\frac{\alpha t_{2}}{2}\left(\begin{array}{cc}
-2 & -\sqrt{3}\\
-\sqrt{3} & 0\end{array}\right)
\end{align*}

With $d_{3z^{2}-r^{2}}$ = $[\begin{array}{cc}
1 & 0\end{array}]$, $d_{x^{2}-y^{2}}$= $[\begin{array}{cc}
0 & 1\end{array}]$, and with a chemical potential $\mu$. We have elected to use a single
parameter $0\leq\alpha\leq1$ to represent the suppression of hopping
in the $z$ direction caused by the superlattice, which is assumed
to effect nearest neighbor and next-nearest neighbor hopping equally.

\begin{figure}
\includegraphics{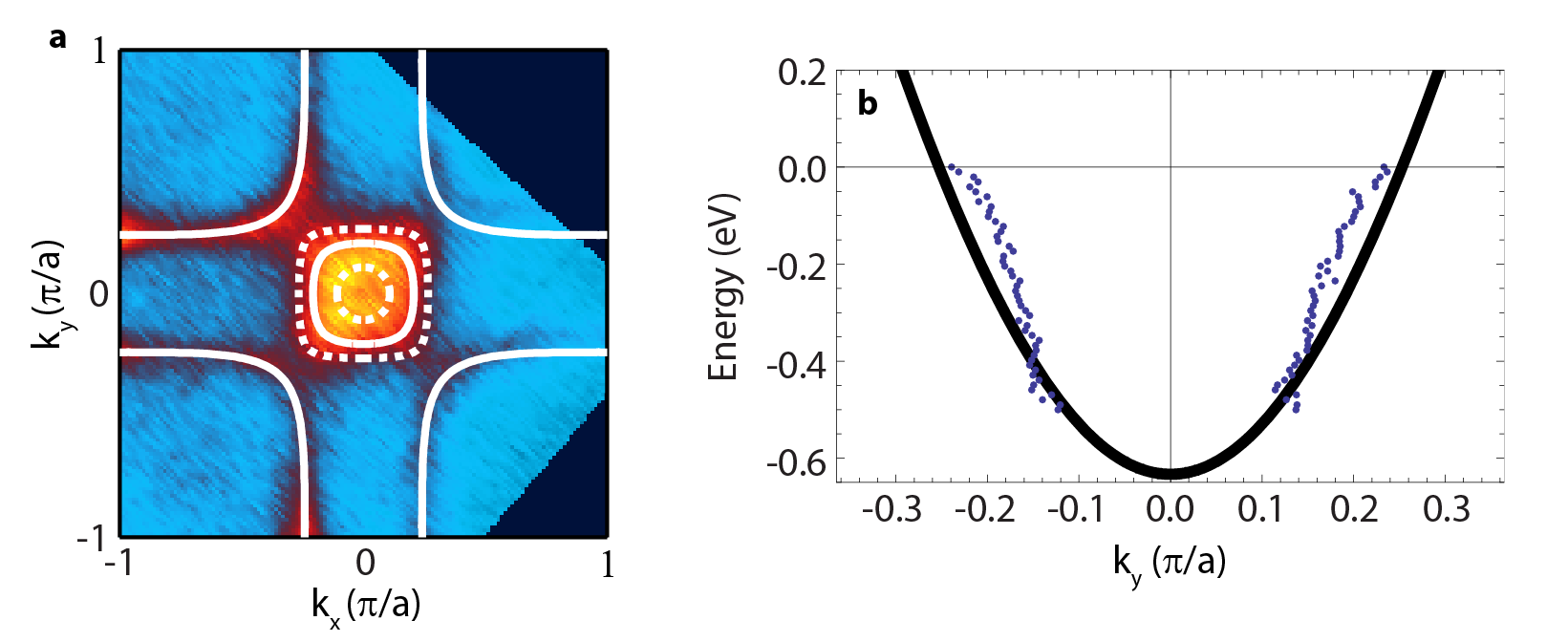}\caption{An example of the tight-binding fit to experimental data, this data
taken from the $n=2$ superlattice. \textbf{a,} The FS overlaid with the hole pocket and electron pocket fits, which are used to determine $t_{2},\mu,\alpha$. The dotted-line shows the upper and lower bounds for the size of the electron pocket. \textbf{b,} The dispersion of the hole pocket at $k_{x} = 0.55 \pi/a$, which is used to determine $t_{1}$. Black lines are
the tight-binding fit. \label{fig:TB_1}}

\end{figure}

This model was fit to our ARPES data for each superlattice and for data from the random alloy La$_{1-x}$Sr$_{x}$MnO$_{3}$ (data not shown) using our determination of the Fermi surface to fit $t_{2}$, $\mu$, and $\alpha$. The procedure for fitting our data is as follows. First, we use the sharply-resolved hole pocket at $E_{F}$ to determine $t_{2}/t_{1}$ and $\mu/t_{1}$. This feature is dominated by segments of the hole pocket near $k_{z} = \pi/a$, which are essentially independent of $\alpha$. With the values of $t_{2}/t_{1}$ and $\mu/t_{1}$ now determined by the hole pocket data, we then determine $\alpha$ by fitting data from the electron pocket, which is very sensitive to the value of $\alpha$. We find that this two-step approach produces much more reliable fit parameters than an unconstrained fit where both Fermi surface sheets are fit simultaneously by allowing $t_{2}/t_{1}$, $\mu/t_{1}$, and $\alpha$ to vary freely. The determination of the size of the electron pocket is the dominant source of uncertainty in determining $\alpha$ and hence the orbital polarization, which is made more complicated by $k_{z}$-smearing that results in an electron pocket with a combination of sharp peaks and a broader background \cite{Krempasky08}. Therefore, we estimate the uncertainty in the size of the electron pocket by taking as an upper bound the FWHM of the intensity around $\Gamma$, and as a lower bound, the separation between two peak maxima around $\Gamma$, as shown in Fig. S\ref{fig:TB_1}a. For the $n = 3$ superlattice, where no electron pocket Fermi surface is resolved, we provide only an upper bound for $\alpha$ that lifts the electron pocket completely above $E_{F}$. We then use the uncertainty in the size of our electron pockets to obtain uncertainty estimates for $\alpha$ shown in Table \ref{tab:TBpar}, also represented in the error bars for the orbital polarization in Fig. 2d. The remaining parameter, $t_{1}$, is then fit to the dispersion of the hole pocket away from $E_{F}$. Note that the value of $t_{1}$ has no effect on the Fermi surfaces in Fig. 2a-c or the numerical values reported in Fig. 2d, since it only results in an overall scaling of the energy units. Our extracted tight-binding parameters for all four samples are displayed in Table \ref{tab:TBpar}.

\begin{table}
\begin{tabular}{|c|c|c|c|c|}
\hline 
Sample & $t_{1}$  & $t_{2}$  & $\mu$  & $\alpha$ \tabularnewline
\hline
\hline 
alloy & 0.87 & 0.13$t_{1}$ & -1.13$t_{1}$ & 1\tabularnewline
\hline 
$n=1$ & 1.1 & 0.15$t_{1}$ & -1.33$t_{1}$ & 0.45$\pm$0.25\tabularnewline
\hline 
$n=2$ & 0.97 & 0.10$t_{1}$ & -1.31$t_{1}$ & 0.22$\pm$0.09\tabularnewline
\hline 
$n=3$ & 1.2 & 0.08$t_{1}$ & -1.38$t_{1}$ & $\leq$0.16\tabularnewline
\hline
\end{tabular}
\caption{The tight-binding parameters that best fit our experimental data. Error bars for $\alpha$ are estimated from the uncertainty in fits of the electron-pocket FS.\label{tab:TBpar}}

\end{table}

\section*{ARPES on LaMnO$_{3}$ and SrMnO$_{3}$}
To ensure that the results reported in the main text are not artifacts from the LaMnO$_{3}$ or SrMnO$_{3}$ surfaces, we have performed ARPES experiments on a series of control samples: a 10 unit cell thick LaMnO$_{3}$ film with MnO$_{2}$ surface termination, an 8 unit cell thick SrMnO$_{3}$ film with SrO termination, an 8 unit cell thick SrMnO$_{3}$ film with MnO$_{2}$ termination, and a 6 unit cell thick SrMnO$_{3}$ film with MnO$_{2}$ termination. To avoid charging effects due to the insulating nature of these films, all samples were grown on 0.5\% Nb-doped SrTiO$_{3}$ substrates and measured at room temperature. Samples were chosen to be thin enough to avoid charging while being thick enough to minimize any signal from the SrTiO$_{3}$ interface.

\begin{figure}
\includegraphics{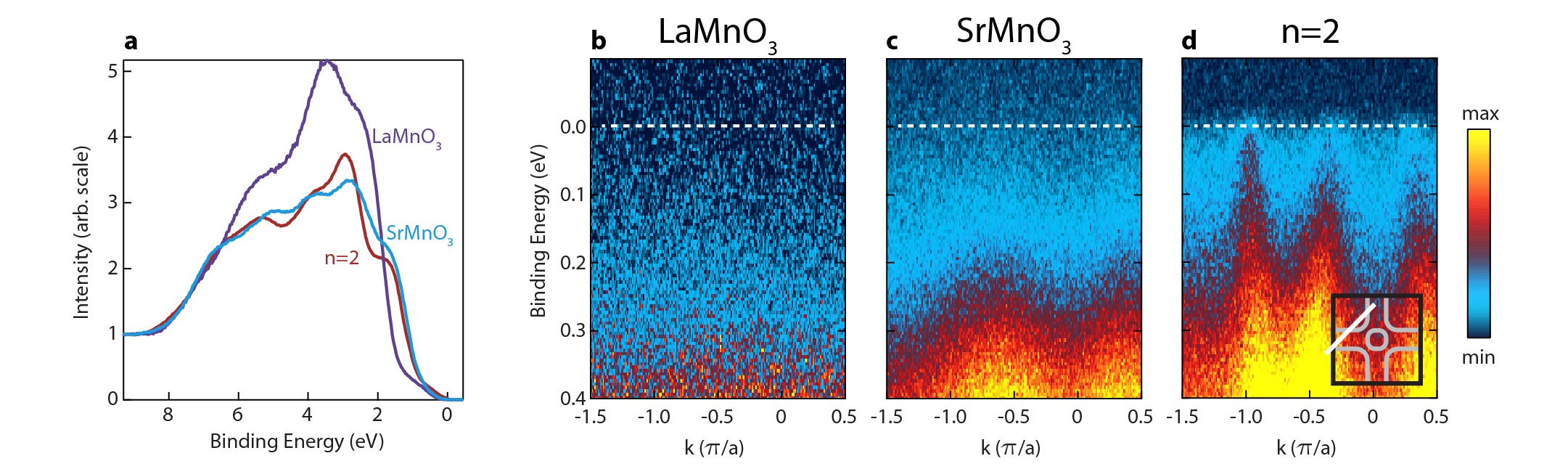}\caption{ARPES from LaMnO$_{3}$ and SrMnO$_{3}$. \textbf{a,} Valence bands of LaMnO$_{3}$ and SrMnO$_{3}$ films (10 and 8 u.c. thick respectively, MnO$_{2}$ terminated), compared with the $n = 2$ superlattice. \textbf{b,c,} ARPES data for the LaMnO$_{3}$ and SrMnO$_{3}$ films showing a lack of dispersive features within 0.4 eV of $E_{F}$. \textbf{d,} ARPES data for the $n = 2$ superlattice showing the dispersive $e_{g}$-derived bands discussed in the main text and responsible for this film's metallic behavior. \label{fig:Control}}
\end{figure}

In all cases, we observe dispersive valence band spectra and sharp LEED patterns, indicative of the high quality of the films. As expected, we did not observe any appreciable or dispersive spectral weight within 0.4 eV of $E_{F}$ for any of the control samples, as shown in Fig. S\ref{fig:Control}. Therefore, we can safely conclude that the dispersive states near $E_{F}$ arise from the LaMnO$_{3}$/SrMnO$_{3}$ interface. The valence band of the MnO$_{2}$ terminated SrMnO$_{3}$ films  qualitatively resemble those of the superlattices at higher binding energies, due to the SrMnO$_{3}$ termination of the superlattices. As expected, we also observe the tail of the SrMnO$_{3}$ valence band (occupied Mn $t_{2g}$ and O $2p$ states) at approximately 0.3 eV binding energy in both the SrMnO$_{3}$ and superlattices (Figs. S\ref{fig:Control}c and d). Nevertheless, only the superlattices exhibit the well-defined, near $E_{F}$ bands which are the focus of our manuscript, confirming that these states arising intrinsically from the LaMnO$_{3}$/SrMnO$_{3}$ interface and cannot be a spurious effect. 

\section*{ARPES background subtraction and $k_{z}$-dispersion}
\begin{figure}
\includegraphics{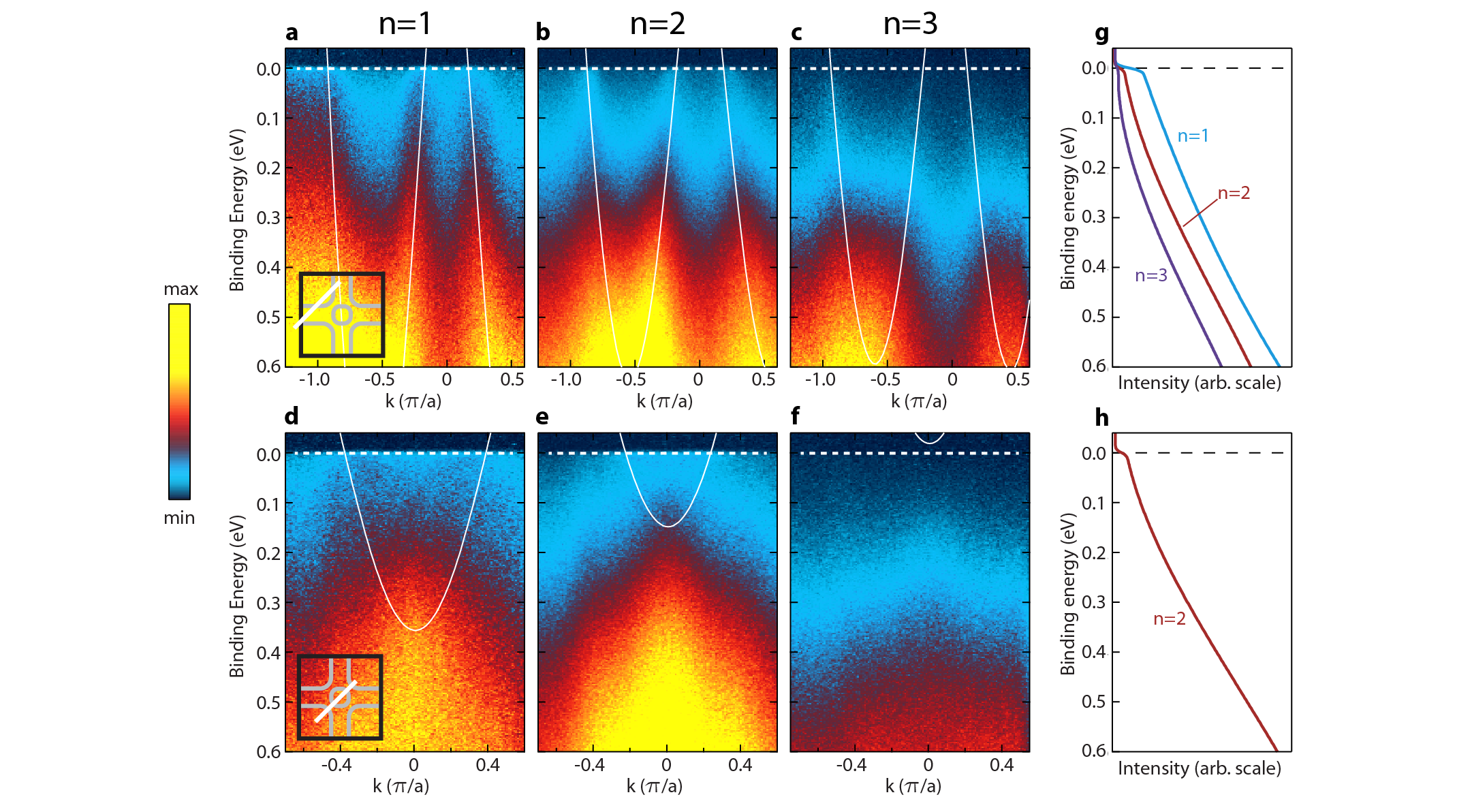}\caption{Raw ARPES data corresponding to the background-subtracted data shown in Figs. 3 and 4 of the main text. \textbf{a-c,} Hole pocket data corresponding to Fig. 3a-c, with TB bands overlaid as white lines. \textbf{d-f,} Electron pocket data for the $n$ = 1, 2, and 3 superlattices respectively, with TB bands for only the electron pockets overlaid as white lines. Panel e corresponds to the data shown in Fig. 4a. \textbf{g,h,} The non-dispersive backgrounds subtracted from the ARPES data shown in the main text. \label{fig:RawARPES}}
\end{figure}

The data presented in Fig. 3a-c and Fig. 4a of the main text have had a non-dispersive background subtracted to more clearly highlight the dispersive bands. In Fig. S\ref{fig:RawARPES} we present the raw data used in these plots, as well as electron-pocket data for the $n$ = 1 and 3 superlattices. Plots and analyses in Fig. 3d,e, Fig. 4b and all other figures use un-subtracted data. The intensity of the ARPES data in Figs. 3-5, and S\ref{fig:RawARPES} were normalized at $12$ eV binding energy. In Fig. 1a-c, the average intensity for the three superlattices are set to be equal at 0.1 eV to facilitate the comparison of the momentum distributions.

The effect of a finite $k_{z}$-dispersion on ARPES spectra of the manganites is well documented for the 3D perovskite \cite{Krempasky08}. In that case, the electron pocket at the BZ center is largely smeared-out due to its significant dispersion along $k_{z}$, while the hole pocket around the BZ corner is expected to show more well defined features due to its largely non-$k_{z}$-dispersive walls. For the superlattices, with a BZ that is a factor of 3$n$ smaller along $k_{z}$, the smearing in the $k_{z}$ direction should extend across the BZ. Thus, in our analysis of the bandstructure we assume that the effects of $k_{z}$ smearing dominate the photoemission, and that any sharply-defined features seen in ARPES correspond to the sections of bandstructure with the least $k_{z}$-dispersion.

